\documentclass{emulateapj}
\usepackage{graphicx}

\usepackage{times}
\usepackage{epsfig}
\usepackage{epsf}
\usepackage{rotating}

\slugcomment{Draft 2009 March 31st, Submitted to PASP}

\shorttitle{$\it GALEX$ cluster variability}
\shortauthors{Browne et al.}

\begin{document}


\title{{\it GALEX} ultraviolet observations of stellar variability in the Hyades and Pleiades clusters}


\author{
Stanley E. Browne,\altaffilmark{1,2},
Barry Y. Welsh,\altaffilmark{1,2} and Jonathan Wheatley\altaffilmark{2}}
\altaffiltext{1}{Eureka Scientific, 2452 Delmer Street, Oakland, CA 94602-3017}
\altaffiltext{2}{
Experimental Astrophysics Group, Space Sciences Laboratory, University of California, 7 Gauss Way, Berkeley, CA 94720. seb@ssl.berkeley.edu}




\begin{abstract}
We 
present $\it GALEX$ near ultraviolet (NUV:1750 - 2750\AA) and far ultraviolet
(FUV: 1350 - 1750\AA) imaging observations of two 1.2$^{\circ}$ diameter
fields in the Hyades and Pleiades open clusters in order to detect possible UV variability
of the member stars.
We have performed a detailed software search for short-term UV flux
variability during these
observations of the $\sim$ 400 sources detected in each of the Hyades and Pleiades fields to
identify flare-like (dMe) stellar
objects. This search resulted in the detection of 16 UV variable sources,
of which 13 can be directly associated with probable M-type stars. The other UV
sources are G-type stars and one newly discovered RR Lyrae star, USNOB1.0 1069-0046050,
of period 0.624 d and distance $\sim$ 4.5 to 7 kpc.
Light curves of photon flux versus time are shown for 7 flare events recorded on six probable dMe stars. UV energies for these flares span the range 2 x 10$^{27}$ to 5 x 10$^{29}$ erg, with a  corresponding variability change of $\Delta$NUV = 1.82 mag. Only one of these flare events (on the star Cl* Melotte 25 LH129) can definitely be associated with an origin on a member the Hyades cluster itself. Finally, many of our M-type candidates show long periods of enhanced UV activity but without the associated rapid increase in flux that is normally associated with a flare event.  However, the total UV energy output during such periods of increased activity is greater than that of many short-term UV flares. These intervals of enhanced low-level UV activity concur with the idea that, even in quiescence, the UV emission from dMe stars may be related to a superposition of many small flare events 
possessing a wide range of energies.
\end{abstract}



\maketitle


\section{Introduction}
M-dwarf stars are the dominant stellar component of our Galaxy by
number, comprising more
than 70$\%$ of the stellar population in the solar neighborhood.
They are almost
fully convective with intense magnetic fields covering most of their stellar disk,
and are thus good astrophysical laboratories for the study of the more extreme
cases of magnetic processes (both coronal $\&$ chromospheric) that occur in
the outer atmospheres of other (more quiescent) stars such as our Sun. 
Recently, their importance
as a major reservoir of stellar hosts for planetary systems has been realized, and
thus estimating the physical conditions and size of a `habitability zone' around such 
UV/X-ray active stars is of 
great value in assessing whether M-star planetary systems are capable of
sustaining living organisms \citep{segura05}. Such a determination is directly
linked to the stellar activity levels and the associated flare energies expected
from each spectral class of M-star. These numbers, unfortunately, are presently poorly known at
both X-ray and UV wavelengths.

Much of our knowledge of the long-term evolution of  stellar coronal activity derives from
X-ray observations of open clusters \citep{stern95}. These stellar associations
constitute a large sample of coeval stars with a similar distance and similar chemical composition, thus providing us with a powerful tool for
a statistical study of the behavior of magnetically active stars.
Comparison studies
show that X-ray emission decreases from younger to older clusters, and this 
decrease in coronal activity with age may be due to rotational spin-down caused by
magnetic breaking. However, the influence of (often unknown) binary companions
in such systems complicates this interpretation, and therefore most X-ray studies have
been observationally biased towards
highly active (and nearby) stars, such that slowly 
rotating stars with relatively weak X-ray activity (i.e. late M-dwarfs)
are generally unaccounted for. 
The majority of X-ray observations often
cannot reveal flux variability on time-scales $<$ 500 sec \citep{pillitteri05},
which biases such results against a flux contribution from
the (more numerous) weakly X-ray flaring dMe stars. This point is
highlighted by the $\it XMM$ observations of Proxima Cen (dM5.5e)
in which varying low-level emission
activity (E $\sim$ 10$^{28}$ erg) was persistent over a period of 45ksec, prior to the
onset of a far larger (and easily more detectable) X-ray
flare event \citep{gudel04}. It has even been suggested that no real quiescence may
be present at all in X-ray coronae, such that the observed emission may be
produced by a superposition of multiple lower energy flares \citep{gudel03}.
Recent multi-wavelength observations
of stellar flares have also revealed an astounding lack of correlation between flares seen
in different wavelength regions  \citep{osten05}, with UV flares
on HR 1099  showing little or no change in the coronal
(X-ray) emission, whereas the UV line-shapes broadened appreciably \citep{ayres01}. 

Over the past 5 years the NASA Galaxy Evolution Explorer ($\it GALEX$) satellite
has been carrying out both
far UV (FUV: 1350 - 1750\AA) and near UV (NUV: 1750 - 2750\AA) broad-band imaging
observations of a large percentage of the sky \citep{mar05}. 
Although designed primarily for UV observations
of low-$\it z$ galaxies, the instrument is proving to be a powerful tool for
the investigation of time-variable and transient stellar phenomena.
For example, long-term ( $>$1500 s)
variability, as determined from
comparisons of orbit-averaged
UV magnitudes, has recently been reported for several hundred objects
in the 
$\it GALEX$ Ultraviolet Variability (GUVV) catalogs
\cite{welsh05,wheat08}. The majority of these UV variable sources
have been shown to be active galaxies, but a non-signifcant fraction of the variable sources
are galactic M-type stars. 
In a more extended search of
$\it GALEX$ data for short-term variability,
49 NUV flare events occurring on time-scales $<$ 150 seconds that
can be associated with
M-dwarf stars of distances
25 to 990 pc and with flare energies ranging from as low as 8 x 10$^{27}$
and up to 1.6 x 10$^{31}$ erg have been detected \citep{welsh07}.
This range of NUV flare energy (and the associated flare light-curve behavior) 
is very similar to that recorded in the visible U-band for nearby (d $<$ 25pc)
M-dwarfs \citep{panagia95}, which argues strongly
for a common emission mechanism for both the
U-band and $\it GALEX$ NUV bands, i.e. that of continuum emission  \citep{hawley03}.
However, for flares recorded in the shorter wavelength FUV band, CIV line emission must
become a major contributor to the observed flux. 

In this Paper we continue our program of UV variability studies using $\it GALEX$
with photometric imaging observations of two pairs of 1.2$^{\circ}$ diameter fields lying within
the Hyades and Pleiades open clusters to investigate the FUV and NUV variability
associated with magnetic activity in member M-dwarf stars. Due to the
intrinsic high UV brightness of some of the stars in each cluster (which can cause saturation
of the $\it GALEX$ detectors), the two pairs of fields were offset by a few degrees from the
nominal centers of each cluster.
Both cluster fields contain several M-stars whose flaring signatures have previously been recorded at
both X-ray and visible wavelengths \citep{stern95, mirz94, hamb97}.
The Hyades cluster
lies at a mean distance of only 46 pc and with an
age of $\sim$ 625 Myr \citep{perry98},
is probably the most well-studied of all open clusters, with
about $\sim$ 300 possible members extending over 20$^{\circ}$ on the sky.
The more distant Pleiades cluster lies at $\sim$ 135 pc \citep{sod05}
and has an estimated age of $\sim$ 110Myr with 
$\sim$ 1400 probable members being listed within the central 6 sq. deg of the cluster \citep{stauf07}.
Although
the average rotation rate for M stars in
the Hyades is only 0.4 times that of the far younger and more active members of the Pleiades
cluster, there is still a large
number of Hyades M-dwarfs that exhibit appreciable chromospheric activity
as revealed by H-$\alpha$ observations
 \citep{terndrup00, stauffer97, reid95}.
X-ray  observations of the Hyades with the
$\it ROSAT$ satellite have
shown that $\sim$ 30$\%$ of its 306 cataloged K and M-type stars are
coronally active down to a limiting X-ray luminosity of 10$^{28}$ erg s$^{-1}$
 \citep{stern95}. Similar X-ray observations of the Pleiades members
resulted in almost all of the catalogued early dMe stars being detected down to a
limiting X-ray luminosity of 3 x 10$^{28}$ erg s$^{-1}$ \citep{mic96}.
All four of our selected fields
for observation with $\it GALEX$ contain
numerous previously identified M-type Hyad and Pleiad members, thus providing us with the potential 
to detect several UV flares emitted from field M-dwarfs during each of
the four $\sim$ 15,000s periods of observation.
In the following sections we describe these observations
and discuss the UV variable sources found in each field and present the
UV flare signatures and associated flare energies for seven identified dMe stars.
 
\section{$\it GALEX$ UV Imaging Observations}
Two sky-fields in the Hyades cluster centered on (A) R.A.04:34:16, Dec.+17:08:56 (2000.) and
(B) R.A.04:27:47, Dec+17:05:24 (2000.) were observed for a total of 15,989 s
and 15,252 s respectively
in both of the FUV and NUV imaging channels of
the $\it GALEX$ satellite \citep{mar05} . Similarly, two sky-fields in the Pleiades cluster
centered on (A) R.A.03:44:20, Dec.+22:00:00 (2000.) and 
(B) R.A.03:45:05, Dec.+26:20:00 (2000.) were observed  for  a total of
33,806 sec and 43,373 s respectively, but solely
in the the $\it GALEX$ NUV channel. In addition, one exposure of
both of the Pleiades fields was recorded in the FUV channel for 1500 s only.
All
these data were recorded as time-tagged photon
events by the $\it GALEX$ micro-channel plate detectors \citep{mor05, mor07}.
Each $\it GALEX$ image has a diameter of $\sim$ 1.24$^{\circ}$ on the sky and
the total observation period for each field was split into 12 (Hyades) and
28 (Pleiades) separate exposures, each of approximately 1500 s duration (i.e. one $\it GALEX$
orbital eclipse period). A few of the exposures were less than the nominal 1500 s duration
due to satellite scheduling and instrumental constraints.
Although most of the exposures were recorded consecutively, the actual observations are
not contiguous due to a 60 minute gap between each exposure during which the $\it GALEX$
detector high voltage is ramped down to avoid the daylight part of the satellite
orbit. All these data were recorded as part of the NASA $\it GALEX$ Guest Investigator
Cycle 2 (ID: GI2-001) and Cycle 4 (ID:GI3-042) programs and the corresponding NUV and FUV images
for each of the 4 sky-fields are shown in Figure 1.  The appearance of these images is now briefly discussed.

\subsection{The UV images of the Hyades Fields (A) and (B)}
These two UV image fields, shown in the upper region of Figure 1, are at a similar
Declination and are almost  adjacent to each other in Right Ascension. The fields
both lie $\sim$ 2.5 $^{\circ}$ from the nominal center
of the cluster, which itself extends over $\sim$ 20$^{\circ}$ on the sky. Due their proximity to the Sun
(d $\sim$ 46 pc), both of theses Hyades UV images
 are essentially unaffected by the effects of interstellar
gas or dust absorption. The brightest UV sources in both of these
$\it GALEX$ images are nearby (d $<$ 10 pc)
F and G-type field stars with a high proper motion. We note that both Hyad (and Pleiad)
M-stars are intrinsically faint, and are generally only observable in
the ultraviolet region due to their enhanced emission
associated with chromospheric and transition region activity.
Based on the previously
known Hyades member stars that are contained within these two fields 
\citep{reid95, stauffer97,
perry98}, there are
4 M-type stars in Hyades-A and one in the Hyades-B field. However,  we note
that these three referenced studies
are magnitude limited ground-based observations in which the intrinsically
faint M-type stars (and those more distant than the Hyades cluster) may not be fully represented. 

\subsection{The UV images of the Pleiades Fields-(A) and -(B)}
These two fields (shown in the lower region of Figure 1) lie approximately $\pm$1$^{\circ}$ in Declination from the nominal center of the
Pleiades cluster, whose stellar membership down to a limiting magnitude of R $\sim$ 20 has been catalogued by
Adams et al. \cite{adams01}. Both of the $\it GALEX$ UV images of the Pleiades
fields appear dramatically
different to those of the Hyades.  Both Pleiad images
reveal substantial nebular UV emission in the form of many filamentary structures
formed by both interstellar scattering and reflection of
radiation from the nearby UV-bright  B-type cluster stars.
Previous UV and IR studies of the Pleiades region by Gibson and Nordsieck \cite{gibson03}
have shown that the majority of the observed UV emission is
from forward scattering of foreground interstellar dust grains.

The brightest stellar sources in both of the Pleiades UV images are,
in common with the Hyades images, dominated by nearby ( d $<$ 10pc) F and G-type stars.
Using the positions of known M-type stars in these two fields, as  listed
by Adams et al. \cite{adams01} and J. Stauffer (private communication),
there are 7 previously catalogued M-type stars in the Pleiades-(A) field and 10 M-type stars in the Pleiades-(B) field.

\section{Data Reduction}
The photon data for each image exposure were processed using Version 5 of the $\it GALEX$ Data Analysis 
Pipeline operated at the Caltech Science Operations Center, Pasadena, CA.,
 \citep{mor05}. The final data product is a 
flat-field corrected photometric time sequence of photons positionally mapped in Right Ascension and 
Declination to the sky. The $\it GALEX$ pipeline then utilizes the SExtractor program
of Bertin $\&$ Arnouts \cite{bertin96} for the detection and photometry of UV sources contained
within each of these NUV and FUV photon image fields.
This procedure produces
a catalog of $\it GALEX$ UV sources, which typically are brighter than NUV magnitude $\sim$
23.0 for a nominal 1500 s observation \citep{mor05}.
However,
due to edge-effects in the $\it GALEX$ images we only used
the data contained within the central 0.55 $^{\circ}$ radius of each field for 
our subsequent
scientific analysis of UV sources. In addition,  the
$\it GALEX$ images are subject to a number of noise-producing signals (such
as reflections from the edge of the detector) that are called `artifacts' and many (but not all) are duly flagged
in the quality assessment phase of data processing which we removed from
our analysis. Also,
the Pleiades images have many diffuse
and extended features due to UV emission from foreground interstellar 
gas and dust, which can confuse the SExtractor source detection algorithm
at the faintest detection levels. Thus, in order
to reduced the number of `false' UV source detections, we limited our search
to sources brighter than NUV magnitude = 22.5 and to those sources with a
measured $\it GALEX$ point spread function of $<$ 8 arcsec fwhm (i.e. stellar and
not extended extragalactic sources).
Our search procedure revealed that
Hyades Field-A contained 335 UV
stellar sources, Hyades Field-B contained 345 UV sources,
Pleiades Field-A contained 323 UV sources and Pleiades Field-B contained 502
UV stellar sources. 

The list of exposure-to-exposure  NUV (and corresponding FUV) magnitudes for each of the
previously identified UVsources, together
with their associated (1-$\sigma$) measurement 
errors, was then queried
to determine potential source variability for a given image field over the time
series of observations. 
Statistically significant stellar variability, as opposed to variations in
the background Poisson noise, was deemed to be real if the largest  difference between the
the set of source magnitude measurements exceeded 2 x 1-$\sigma$ measurement error (i.e.
typically $>$ 0.3 mag). This initial variability selection criterion was based on extensive searches of the $\it GALEX$ archive that
resulted in the assembly of the two GUVV catalogs
\cite{welsh05,wheat08}. Unfortunately, due to the presence of the many artifacts (listed previously) that 
affected the present $\it GALEX$ images, the sole use of this
statistical test in revealing low levels of source variability often produced many false positive detections.
Hence, actual verification of the true long term variable nature of these sources required individual
visual inspection of their stellar images (to reveal low count rate artifacts), in addition to a short statistical comparison
of the source photon list data for each exposure. This latter statistical test on the photon data involved a comparison
between the mean source count rate level (and its associated
standard error) established for each 750s of every exposure (i.e. half an observation period). Stellar
count rate variability
was deemed significant in a similar manner to that established for the exposure-to-exposure magnitudes, in that
the count rate level needed to exceed 2 x 1-$\sigma$ of the count rate standard error.

The application of these two selection criteria resulted in the identification of a total of
6 variable stellar Hyad candidates and 8 variable stellar Pleiad candidates. The
UV variable sources found in each of the 4 observed fields are listed in Table 1 together with their galactic co-ordinates and their respective observed maximum and minimum FUV and NUV magnitudes
(FUV$_{max/min}$, NUV$_{max/min}$) . In cases where these sources
have previous identifications in the Simbad database and/or they are listed
in either the USNO-B1.0  \citep{monet03} or
the Two Micron All Sky Survey  (2MASS) catalog \citep{cutri03}, these are also listed in Table 1 together with their 2MASS (H - K) and (J - H) color magnitudes.

As a check on finding (short-term) flaring and/or variable
sources whose UV output may have changed over time intervals much shorter than one $\it GALEX$
exposure (i.e. $<$$<$ 1500 s), or with a small
flux change that, over the integration of one exposure, may have
been undetected by our exposure-to-exposure
magnitude comparison method, we subsequently inspected the time-tagged photon list files for all of
the UV sources present in each of the 4 image fields. Since this involves the inspection of
a very large amount of data, we used a crude data compression method to
inspect the photon data for all of the UV sources contained within each individual exposure.
We used the
 `varpix'  variability search algorithm
\citep{welsh07}, in which the
software tool bins all of  the photon data 
accumulated in consecutive 16 s time intervals over  image areas of 12 arcsec$^{2}$ pixels for each exposure.
Intensity variability for each of the these `super-pixels' as a function
of time throughout an exposure period is then assessed
against the median and maximum photon flux value
to determine a `variability signal-to-noise ratio' for each of these large image pixels. We identiified
potential variable sources in our present images as those with a variability signal-to-noise ratio $>$  8:1.
This ratio was chosen through a trial and error approach on several $\it GALEX$ images
 in order to minimize the many false positives that occur
at lower variability S/N ratios. These false variability detections at low S/N are caused by the flux from bright objects
spilling over into adjacent `super-pixels', thus causing the appearance of source variability.
For S/N variability $>$ 8:1, we deemed that if
the variation in source flux was due to
a flaring dMe star, then a characteristic flare light curve
should emerge from an inspection of the `varpix' output \citep{welsh07}. The `varpix' search method confirmed all of
our 14 previously detected variable sources, with the addition of
one new flaring source, SDSS J0422842.78+171149.6, 
within the Hyades-B field.

For sources with only one detection over the entire series of exposures (i.e. transients), 
the previous statistical
tests for variability were not applicable. We therefore set a criterion for the examination of all single detections of UV sources
brighter than a limiting magnitude of NUV $<$ 23.0 We then examined the photon list light curves
for all of these selected sources using the `varpix' software tool, which resulted in the discovery
of one new source, Hyades-A 04:33:56.6 +16:52:09.6.
From the application of all of these search methods we believe that no single UV (long and short-term) variable star
brighter than NUV magnitude $\sim$ 22.5 was missed in our present variability search, the results of
which are listed in Table 1.

\section{Results and Discussion}

\subsection{The Identification of M-type Stars}
The particular topic of present interest is the enhanced levels of chromospheric
and coronal activity on dMe stars that can produce large stellar flares which are observable at
ultraviolet wavelengths. Therefore, we
need to determine which of the UV variable objects listed in Table 1 can be directly
associated with dMe flare stars. 
We note that
most ground-based visual studies of stars 
seen towards both the Hyades and Pleiades clusters
are magnitude limited observations in which the
intrinsically faint M-type stars may not be fully represented
\citep{perry98, dobbie02,
stauf07}.
Therefore, previous
estimates of both the numbers of M-type stars and their possible membership
status of both clusters are far from being complete. In addition, the fraction of 
chromospherically active M stars peaks
at spectral type M7 \citep{west08}, such stars often being too faint to be detected in 
many visible studies.
Thus, without deep
multi-band visible photometry and a measurement of stellar proper motion and
stellar spectral type, obtaining accurate 
estimates of the number of
all possible UV active M-type stars in either of our cluster fields (and their cluster membership status)
is beyond the present scope of this paper. 

Fortunately 2MASS photometric stellar magnitude data (as listed in Table 1) are available
for both of our observed regions \citep{cutri03}, whereas Sloan Digital Sky Survey (SDSS) data are
only available for a small area of one of the Hyades-B  fields \citep{york00}.
Relationships between the SDSS and 2MASS color
magnitudes have been derived for $\sim$ 38,000 low-mass stars \citep{west08}, and
can be used to identify possible M-type stars. In Figure 2 we show the locus of
a 2MASS (H - K) versus (J - H) color-color diagram
that encompasses M0 to M9 spectral types \citep{west08}. We also show
the positions of the UV variable
sources listed in Table 1. This figure clearly shows that 12 of our sample of UV variables lie in
the region of the plot where M-type stars are to be expected to be found.  Three of the 4 outlying
objects  lie to the left of the main grouping of Figure 2 and are
thought not to be M-type stars. These non M-type stars are now
briefly discussed in Section 4.2 prior to a more detailed discussion of the remaining 13
sources in Section 4.3, which we argue are probable M-type stars. 

\subsection{Non M-type Hyades and Pleiades UV Variables}

\subsubsection{Hyades-B source 04:27:53.6 +16:51:36.0}
The `varpix' light-curve for this object showed no sign of short-term variability that could
be associated with flaring, but instead
revealed the source to possess a near-constant level of increased emission measured during two
exposures compared
to that observed over other observation periods.  Its 2MASS colors are more consistent with
a star of spectral type F or G  \citep{fin00}, and we note that $\it ROSAT$ observations of
the Hyades revealed a high detection rate for known G-type stars and binary systems \citep{stern95}.
Thus, the cause of the weak (and long duration) UV variability
that we have observed for this object could
possibly be due to the presence of a close companion star. 

\subsubsection{Hyades-B source 04:28:32.4 +16:58:21.6}
Figure 3 shows the NUV light-curve for this source, which shows clear
periodic variability with a maximum increase of  $\Delta$NUV = 1.64 mag. and
$\Delta$FUV $>$ 2.76 mag.
This type of UV flux variation is very similar to that observed by $\it GALEX$ for RR Lyrae stars
\citep{wheat05}), in which the stellar brightness variation is primarily due to radial pulsations that produce
an observed temperature change that is mostly pronounced in the UV.
Under the assumption of its RR Lyrae nature, we have derived a
period of 0.624 d for this source, which is a typical value for this type of star.

Using a 2MASS K magnitude of 13.5,  E(B-V) = 0.38 and the period-luminosity relation of
Sollima et al. \citep{sol08} for RR Lyraes, we derive a distance of $\sim$ 4.5 to 7 kpc for this star,
placing it well out into the galactic halo.

\subsubsection{Pleiades-B source 03:42:59.0 +26:17:01.0}
The `varpix' light-curve for this source revealed an increased level of  UV emission during
only a few oribits,
with no obvious associated short-term flare signature. Its 2MASS colors are consistent with a star of
spectral type earlier than G5 \citep{fin00}, and $\it ROSAT$ X-ray observations of the central region of
the Pleiades cluster revealed a high detection rate for dwarf G-type stars \citep{mic96}.
Since binary dG stars are more intense X-ray emitters than single dG-type stars, it seems 
 probable that binarity may well be
the cause of its observed UV variability.

\subsection{M-type Hyades and Pleiades UV Variables}
The following section discusses likely M-type stars that were detected as UV variables
in both the Pleiades and Hyades fields. For the 6
stars that exhibited flaring signatures, we show their NUV light-curves (i.e. counts versus time) in Figure 4. Note that Pleiades-B source 03:43:35.5 +26:21:31.1 was observed flaring on two occasions. All flaring signatures were also detected in the FUV channel for both Hyades fields, but the
photon data are of
a much lower signal-to-noise ratio and are therefore not shown in Figure 4.

The UV light-curves of a sample of $\sim$ 50 dMe flare stars observed
with $\it GALEX$ have been previously investigated \citep{welsh07}. The
authors found 3 distinctive signatures of these flare events in plots
of photon flux versus time (see Figure 2 of their paper),
and the present NUV flare light-curves shown in our Figure 4
are all qualitatively consistent with those flare signatures.

Finally, we remind the reader that
$\it GALEX$ has an uneven observational cadence
with at
least a 60 minute gap between individual
exposures. On certain occasions (due to satellite operational constraints) this
gap was several hours long. Thus, although $\it GALEX$ is an ideal detector of UV emission from both large
and small flare events, it is not an ideal tracer of the time evolution of flares for periods
longer than 1500 s. As such, these observations are not  well suited for accurate determinations
of flare activity rates recorded over time periods $>$ 1500 s. In the following subsections we briefly discuss the UV variability detected on the M-type
stars in both cluster fields.

\subsubsection{Hyades-A source 04:33:56.6 +16:52:09.6}
This was the only transient source detected by our variability search.  It showed no conclusive
UV flare signature in its `varpix' light curve during the single exposure in which
it was detected and it is listed with a spectral type of M1 in Simbad.
It has been classified as a Hyades cluster member through
proper motion studies \citep{reid92}, and has also been detected at X-ray wavelengths
\citep{stern95}. It is likely that our detection of this source in the UV was caused by
observing it after a large stellar flare which had yet to return to its low pre-flare
flux level (which in this case was beneath the detection limit of
$\it GALEX$). Another plausible explanation for the transient nature of this source could
be the detection of UV emission from an eruptive binary companion star. 

\subsubsection{Hyades-A  source 04:34:31.3 +17:22:20.1}
This source showed no flaring signature in its `varpix' light curves and its 2MASS colors
are consistent with an M-type spectral classification. It has
no previous history of flare activity, but was detected by
$\it GALEX$ in all of the 12
NUV exposures with the majority of detections lying in the 19.9 $< $ NUV $< $20.3 mag range.
Two of these detections were $\sim$ 0.25 mag. brighter in the NUV. 

In Figure 5 we show a concatenated series of 10
exposures (in the form of light-curves of NUV photon flux versus time) for this source. 
This plot represents the light-curves from all of the exposures strung together, and
does not represent a contiguous time-series of observations. However,
it is apparent that the overall flux level recorded over a total of  $\sim$ 15,000 s of observations
(but in reality recorded over an actual period $\sim$ 24 hours) is of a near-constant nature.
We note that towards the end of the observational period there was a slow, but steady increase
in activity (of $\sim$ 30$\%$) for this source that was below the threshold of both of our variability
detection techniques.
The enhanced activity level is long-lived
($>>$ 3000 s) and thus when viewed over one exposure period the
light-curve signature appears to be of a `quasi-constant' nature. We
believe that this type of light-curve is best explained as an extended period of repeated low-level
flare activity following the short
emission period of a large flare event (which unfortunately our $\it GALEX$ observations missed).
We shall return to the importance of this form of elevated activity in Section 4.4.

\subsubsection{Hyades-B source 04:26:04.4 +17:07:14.0}
This source was recognized as
a flare star in the $\it ROSAT$ X-ray survey of the Hyades \citep{stern95}.
Its membership of the
Hyades cluster has been confirmed by Leggett et al. \cite{leg94} who derive a distance
of 46.8pc and they classify the star as a possible M-dwarf binary. Its associated 
NUV light curve is shown in Figure 4 for the flare event presently
observed by $\it GALEX$.  This event has
a fast rise-time ($\sim$ 20 sec) followed by a `quasi-exponential' decay, which also
exhibits a secondary emission peak at time = 480 s. We note that the flare
on this star was the largest of all 7 flare events observed with $\it GALEX$ by an order of
magnitude.
This star was also observed
in a second flare outburst, but with an intensity far smaller than the major event.
This star has SDSS DR6.0 photometric colors indices of (r - i) = 1.38 and (i - z) = 1.22, which would suggest a spectral type of M3.5 to M6.5 \citep{west08}.

In Figure 6 we show a more detailed plot of the NUV and associated FUV light-curves 
recorded by $\it GALEX$ during the one exposure in which the large flare event occurred.
It is clear that the FUV channel follows the same light-curve signature as the NUV channel.
Within the measurement error of $\it GALEX$ the start times of both the primary (t $\sim$ 180 s) and
secondary (t $\sim$ 490 s) flare events are the same.

We have also plotted  the ratio of FUV to NUV flux versus time for the same exposure period (i.e. 
a color ratio plot) in Figure 6.
The photon count rates have been converted to fluxes (in erg cm$^{-2}$ s$^{-1}$ \AA$^{-1}$)
using the appropriate conversion factors 
for the $\it GALEX$ instrument \citep{mor05}.  This plot can be directly
compared to the one shown in Figure 1 of Robinson et al. \cite{robinson05} for the giant
flare recorded by $\it GALEX$ on the dM4e star GJ 3685A. In both cases the FUV/NUV ratio
increases rapidly to values greater than unity, which is then followed by a period of exponential
decay in which the NUV flux exceeds that of the FUV. Finally the flux ratio value rises 
above unity again as a second flare evolves (at t $\sim$ 490 s). The classic explanation for such
behavior is that FUV line emission (from CIV $\lambda$1550\AA) dominates the early stages of
flare evolution, which is subsequently followed by dominance by NUV line and/or continuum
emission.

\subsubsection{Hyades-B source 04:27:33.6 +16:52:22.0}
This source is identified as the star Cl* Melotte 25 LH 110, with
reported photometric magnitudes of  B=16.9, V=15.29, R=14.15 and J=10.9
by Leggett $\&$ Hawkins \cite{leg88}. These values
are consistent with an M-type spectral classification for this star, as are its 2MASS colors
listed in Table 1. The source also has SDSS DR6.0 photometry with color indices of (r - i) = 1.15
and (i - z) = 1.12, which would indicate a spectral type of M2.5 - M5.5  \citep{west08}.
The trigonometric
distance for this M-type star is 26 pc,
which would place it at the very periphery of the Hyades cluster  whose
stellar membership is thought to span the 25 - 65 pc distance range \citep{perry98}.
This star has not been listed as a possible Hyad member,
since its proper motions of 2 mas yr$^{-1}$ (RA) and -14 mas yr$^{-1}$ (Dec) are
inconsistent with Hyad cluster membership \citep{reid92}.

In Figure 4
we show the photon flux as a function of time (recorded over one exposure period) for the
observed flare on this source. To reveal the unusually active nature of this target,  we also show
the 
concatenated light-curves for all 10 exposures of this star in Figure 5. 
We see that in addition to the large flare at t $\sim$ 12000 s there is
also a far smaller flare event occurring at t $\sim$ 5500 s. In addition,
there is a period of
enhanced UV activity starting at t $\sim$ 13700 s. which is similar in nature to
that discussed previously for the source Hyades-A 04:34:31.4 +17:22:20.1.

\subsubsection{Hyades-B source 04:28:42.7 +17:11:50.3}
The SDSS DR6.0 color indices for this star of (r - i) = 1.71 and (i - z) = 1.01 suggest a spectral
type of M4 to M5 \citep{west08}. There is no additional catalogue information for this star and thus we are unable to speculate
whether it is a Hyad member and we therefore place a conservative distance estimate
of  20 - 50 pc for this star.

The UV flare signature for this source shown in Figure 4 is very weak, and although of low S/N ratio, it appears
to be quite a long-lived event with an extended
period of
activity lasting $>$ 300 s.

\subsubsection{Pleiades-A source 03:42:35.6 +21:50:31.0}
This UV variable source is the star V614 Tau, which has a history of previous optical flare
activity \citep{haro82}. Its NUV light-curve shown in Figure 4 is probably the
most complex of all of the flares we have observed. Although of
low S/N ratio, the data reveal two flare
intensity peaks separated by $\sim$ 40 s followed by a very extended period of diminishing activity
that lasts at least 300 s. The rise time for the initial flare is $>$ 50 s, which
is unusually long compared with other UV flare vents we have detected.
We also note the statistically significant small `bump'  that occurs
$\sim$ 40 s prior to the onset of the first main flare event.  Pre-cursor flares of this type
have routinely been recorded on the Sun at both visible and X-ray wavelengths.
A cluster membership probability has not been assigned
to this star \citep{stauf91}, and thus we place a conservative distance estimate for it of between 20 - 130 pc.

\subsubsection{Pleiades-B source 03:43:35.5 +26:21:31.1}
This variable source is the high proper motion star NLTT 11679 (173 mas yr$^{-1}$).
Such a high value of proper motion value rules out its possible membership of the Pleiades cluster  \citep{hamb91}.
The NUV light curve (Flare 1) in Figure 4 is of a classic UV flare signature, with a
fast rise time and `quasi-exponential' decay that lasts $\sim$ 150 s  \citep{welsh07}.
However, the second flare observed on this star (Flare 2, Figure 4) has a significantly different
UV light-curve signature with a significantly extended period of activity following the main rise in flux.
These two flare events occurred in consecutive orbits ($\sim$ 60 min apart), with the
smaller flare event (Flare 2) being a pre-cursor to the larger flare. Immediately prior to
the exposure that contained Flare 2, the star was in a quiescent state with NUV$_{mag}$ = 21.7. We place a 
conservative distance estimate for this M-type star of between 20 - 50 pc.

\subsubsection{Pleiades-B source 03:44:26.4 +26:02:31.0}
This is the dMe flare star named MZ Tau which has been catalogued by Deacon
$\&$ Hambly \cite{deac04}
as a member of the Pleiades cluster based on proper motion studies. 
It has been observed to flare at visual wavelengths  \citep{chav75},
whereas our the `varpix'  plot of our UV data did not show any flaring signature. Instead, its
UV variability was observed as a gradual increase in flux level during one exposure, with the majority of
the remaining
observations being at the lowest limit of our detectability. Immediately prior to the brightest exposure of
NUV$_{mag}$ = 20.72, the two preceding exposures were of NUV$_{mag}$ = 21.6 and 21.9.
This flux variability behavior suggests that MZ Tau was in an increasing state of activity, presumably
prior to a large flare event whose peak flux was missed by our observations.

\subsubsection{Pleiades-B source 03:45:03.8 +26:11:08.1}
This source was an outlier from the main group of probable M-stars shown in Figure 2,
but had large errors on its 2MASS color indices. Also it was positioned to the right
of the main M star grouping, as opposed to the other 3 outliers which were
positioned to the left of the main group. We note that 
the USNO-B1.0 image of this source suggests that it may have a companion star, which
may explain the anomalous 2MASS color indices.
It remained beneath detection levels for all but 5 of the NUV exposures, and although
no flare signature was observed in the `varpix' light curve for this object, there was
the sizable increase of $\Delta$NUV = 2.17 mag
over its lowest observed magnitude (i.e. the largest magnitude
variation in all of the Pleiades observations).
There is no catalogued information for this star and thus we cannot assess whether
it is a cluster member or not.

\subsubsection{Pleiades-B source 03:45:43.6 +26:05:05.0}
This star (Cl* Melotte 22 SK 507) is listed as a non-member of the Pleiades cluster \citep{stauf91},
based on its high proper motion value. Its
2MASS colors are consistent with an M-type classification, and
in Figure 4 its NUV
light-curve shows a short (t $\sim$ 25s) rise-time flare event 
followed by an extended diminishing activity period lasting $\sim$ 300 s. 
Although the data is of low S/N, this diminishing activity period seems to contain
several small flare events occurring after the main flare event. No other
data exists for this star, and
we conservatively place a distance range of 20 - 80 pc for this dMe star.

\subsection{Flare Energies}
Estimates for the total NUV energy emitted from each of the 7 flare events listed previously are
shown in columns 2 and 3 of Table 2. These estimates have been derived from 
subtracting the average of the integrated flux 200 s prior to the onset of the flare and then
integrating the total emitted
flux (shown in Figure 4) over the time period of the UV flare event. Unfortunately accurate distances are 
not available for all of the 6 stars that flared, and in Table 2 we place maximum and
minimum estimates for these energies based on the distances given in the previous Section.
For flare stars with known distances, the values in columns
2 and 3 are identical.

The flare energy values found for both clusters are all in the 2 x 10$^{27}$ to 5 x 10$^{29}$ erg range, which (on average)
are $\sim$ two orders of magnitude lower than that found in a $\it GALEX$ survey of M star flares
\citep{welsh07}. This level of flare energy is similar to that
found in varying low-level emission activity 
in X-ray observations of the nearby dM5.5e flare star Proxima Cen \citep{gudel04}.
Our detection of such low energy events can either be due to only small flares being produced on the
dMe stars in both cluster fields, or that the Welsh et al. study was biased towards the detection of
far more energetic events. We favor the latter interpretation, since
in the Welsh et al. survey of UV flare events found in 
the (then) available $\it GALEX$ data
archive, they found an average change of $\Delta$NUV = 2.7 mag.
for events on suspected dMe stars. If we presently restrict our selection of flare events
to the 6 stars shown in Figure 4, then 
inspection of their NUV magnitudes listed in Table 1 reveals 
an average magnitude change of $\Delta$NUV = 1.82 mag.
We note that the $\it GALEX$ survey study of M-dwarf
variability was performed using the `varpix' software
search set for a variability S/N ratio of $>$ 15:1 \citep{welsh07}, thus biasing the detection of
larger variable events than those of our present study. For the case of the 3 flare
variable events found in the Hyades fields
we derive an average magnitude change of $\Delta$FUV $>$1.39 mag.

In Section 4.3.2 we noted that there was at least a $>$ 3000s period of enhanced UV activity
on the Hyades-A  source 04:34:31.3 +17:22:20.1.  If we integrate the excess flux above
that of the background level (shown as a dotted line in Figure 5) over one exposure period,
we derive a total energy of 5.5 x 10$^{27}$ to 3.4 x 10$^{28}$ erg (assuming a minimum
and maximum distance to the source
of 20 - 50 pc). We note that this increase in energy is far greater than that attributed
to the short period flare shown in the upper plot of Figure 5. This result is significant, since
many of the M-type variables were observed in this `quasi-constant' state of increased
flux in addition to being observed in a classic flaring mode. Since these periods
of increased activity last for significant time intervals, they would seem to be the major
contributer of UV energy output for dMe stars. These observations thus raise the intriguing question as to which type of
increasing flux versus time signature actually constitutes recognition as a
flare event  and which is the more important with regard to total energy output?
Our present data would suggest that the
periods of increased NUV and FUV activity that have no accompanying classic flare signature 
may contribute a far larger energy output, and this discussed in more detail in Section 4.5.
 
\subsection{Discussion}
Unfortunately, 
our $\it GALEX$ observations of both cluster fields have revealed only one
dMe flare star (Hyades-B source 04:26:04.4 +17:07:14.0) as being a definite cluster member.
Thus, we are presently  unable to carry out a meaningful comparison of the UV activity of
dMe star members in both clusters. Additionally,
due to the uneven cadence observations made by the $\it GALEX$ satellite, we believe
that although the UV wavelength region is clearly ideally suited to the detection and observation
of flares occurring on known dMe stars, the visible regime is currently better suited to assess
chromospheric activity rates. Such optical observations should be carried out on large numbers
of cluster members (whose proper motion, spectral type and distance have previously been
determined) using sensitive
H-alpha spectral measurements observed over more
extended time periods \citep{reid95, stauffer97}.
Furthermore, our UV observations have raised the question as to which type of
increasing flux versus time signature actually constitutes recognition as a true
flare event and which is the more important with regard to total UV energy output.
For example, two sets of UV light-curves shown in Figure 5 may well be both associated
with dMe stars but only one of the stars (Hyades-B 04:27:33.6 +16:52:22.2) can definitely be
confirmed as being a flare star that exhibits a classic flare signature. The $\it GALEX$ 
observations of the other star
(Hyades-A 04:34:31.3 +17:22:20.1) revealed only
periods of increased NUV and FUV activity with no actual (short-term) flare signature being recorded. However,
the total energy output over one exposure period of increased
UV activity was greater than that attributed to many classic short period flares.
This latter observation concurs with  X-ray data for other
flare stars that
suggest that there may not be a true quiescent state for dMe stars \citep{gudel03}.
Instead, the observed
low energy state may actually be a superposition of many small, but long lasting, flare events.
Both X-ray and visible time-resolved spectra of 
flare events \citep{gudel04, fuhr08}) both show line and
continuum emission variations that could generate the multiple peaks and substructure
that we have observed in both our NUV and FUV light-curves. Clearly theorists need to
fully explain which physical processes in the chromospheres and coronae on
dMe stars that give rise to all of this substructure observed at different wavelengths. This 
may be relevant in explaining the difference in the shorter rise time of flares observed
at visible wavelengths (t $\sim$ 10 s) compared with those presently
detected at UV wavelengths (t $\sim$ 50 s).

\section{Conclusion}
We have presented  a time-series of near and far ultraviolet imaging observations of four 1.2$^{\circ}$ diameter fields along sight-lines to the Hyades and Pleiades open clusters using
the $\it GALEX$ satellite to investigate
possible UV variability of the stellar members. Stellar UV sources in each cluster field were
extracted from each exposure image recorded over a total observing
period of $\sim$ 15,000 s,  and their corresponding NUV and UV source magnitudes derived.
These exposure-to-exposure source magnitudes were then queried to reveal possible
UV variability over the time-series of observations. In addition other time variability tests were carried
out on the actual photon list data. These search methods revealed 16 UV variable sources, whose
maximum and minimum variations in NUV and FUV source magnitudes are listed in Table 1.

The UV images of all 4 fields are dominated by the presence of bright (nearby) F and G field stars,
with the Pleiades images showing UV emission in the form of filamentary structures due to
the scattering and refection of UV starlight by interstellar gas and dust grains. 

We have used a 2MASS color-color indices plot to identify possible M-type stars from our
list of 16 UV variables. This method revealed two G-type stars and one
previously unknown RR Lyrae star
(Hyades-B source 04:28:32.4 +16:58:21.6, with a derived period of 0.624 d and a distance of $\sim$ 4.5 to 7 kpc), that were
clearly separate from the remaining 13 variables whose spectral types were consistent
with that of M stars. Of these 13 possible M-type stars we have detected 7 stellar flare events recorded
towards 6 probable dMe stars. Light-curves (flux versus time) are presented for these 7 events.
The majority of these M-type UV variable stars did not exhibit a `classic'  flare signature in plots
of their UV flux versus time.  Instead, these sources were observed during periods of near-constant but elevated levels of activity compared with exposure periods in which only lower flux levels
were detected.

Energies for the 7 flare events have been derived from the
photon flux versus time plots, using estimates of the distances to these sources. The maximum and
minimum flare energies (based on the distance uncertainties) span the range 2 x 10$^{27}$ to 5 x 10$^{29}$ erg, which is about
2 orders of magnitude less than the average flare energy found in a survey of $\sim$
50 dMe stellar flares
observed with $\it GALEX$  \citep{welsh07}.
This anomaly can be explained by the fact that in the latter study the software detection methods were
biased towards the discovery of far larger flare events. In our present study of the two
open clusters we find an average variability change of $\Delta$NUV = 1.82 mag for the six 
dMe flare stars over the 15,000 sec of observations. Rather surprisingly, only one of the flare events 
could be definitely associated with an outburst on a star
that is a Hyades member (i.e. Hyades-B source 04:26:04.4 +17:07:14.0), with all the other 
flare events occurring on stars with indefinite cluster membership.

\begin{acknowledgements}
We gratefully acknowledge NASA's support for construction, operation and science analysis for the GALEX mission,
developed in cooperation with the Centre National d'Etudes Spatiales
of France and the Korean Ministry of
Science and Technology. We acknowledge the dedicated
team of engineers, technicians, and administrative staff from JPL/Caltech,
Orbital Sciences Corporation, University
of California, Berkeley, Laboratoire d'Astrophysique de Marseille,
and the other institutions who made this mission possible.
We also thank Suzanne Hawley and John Bochanski (University of
Washington) and Andrew West (UC Berkeley) who gave excellent
guidance and advice in writing this paper. 

Financial support for this research was provided by the
NASA $\it GALEX$ Guest Investigator program, administered by
the Goddard Spaceflight Center in Greenbelt, Maryland.
This publication makes use of data products from the SIMBAD database,
operated at CDS, Strasbourg, France. 
\end{acknowledgements}

\newpage
\begin{table*}
\begin{center}
\caption{UV variable sources}
\scriptsize
\begin{tabular}{cllcclllcc}
\hline
\hline
$\it GALEX$ Field&R.A.(J2000)&Dec.(J2000)& FUV$_{max}$&FUV$_{min}$&NUV$_{max}$&NUV$_{min}$&Simbad Identification&(H - K)&(J - H)\\
\hline
\hline
Hyades-A&04:33:56.6&+16:52:09.6&22.71&$>$23.5&21.31&$>$22.5&Cl* Melotte 25 Reid 332&0.630&0.267 \\
Hyades-A&04:34:31.3&+17:22:20.1&20.47&20.78  &19.66&20.37&USNOB1.0 1073-0063497&0.265&0.706\\
Hyades-B&04:26:04.4&+17:07:14.0&18.14&20.47  &17.56&21.67&Cl* Melotte 25 LH 129&0.289&0.611 \\
Hyades-B&04:27:33.6&+16:52:22.2&21.38&22.17  &20.18&20.91&Cl* Melotte 25 LH 110&0.225&0.639 \\
Hyades-B&04:27:41.2&+16:33:09.4&21.61&$>$23.5&20.84&22.46&IRXS J042738.6+171837&0.573&0.268 \\
Hyades-B&04:27:53.6&+16:51:36.0&22.98&23.36  &20.49&21.17&USNOB1.0 1068-0045793&0.184&0.302 \\
Hyades-B&04:28:32.4&+16:58:21.6&20.74&$>$23.5&19.30&20.94&USNOB1.0 1069-0046050&0.119&0.307 \\
Hyades-B&04:28:42.7&+17:11:50.3&22.44&$>$23.5&21.39&22.50&SDSS J0422842.78+171149.6&0.323&0.666\\

\hline
Pleiades-A&03:42:35.6&+21:50:31.0&N/A&N/A&20.67&21.98&V614 Tau&0.188&0.672 \\
Pleiades-A&03:42:36.9&+22:12:31.0&N/A&N/A&21.10&22.49&Cl* Melotte 22 LLP 137&0.337&0.570 \\
Pleiades-B&03:42:59.0&+26:17:01.0&N/A&N/A&20.43&21.54&USNOB1.0 1162-0044156&0.341&0.113 \\
Pleiades-B&03:43:35.5&+26:21:31.1&N/A&N/A&20.08&21.99&NLTT 11679&0.252&0.617 \\
Pleiades-B&03:44:26.4&+26:02:31.0&N/A&N/A&20.72&21.99&MZ Tau&0.155&0.657 \\
Pleiades-B&03:45:29.9&+26:26:12.0&N/A&N/A&20.76&21.75&USNOB1.0 1164-0045615&0.285&0.617\\
Pleiades-B&03:45:03.8&+26:11:08.1&N/A&N/A&20.24&22.41&2MASS 03450387+2611053&0.207&0.844\\
Pleiades-B&03:45:43.6&+26:05:05.0&N/A&N/A&20.61&22.37&Cl* Melotte 22 SK 507&0.277&0.555 \\
\hline
\hline
\end{tabular}
\end{center}
\end{table*}
\begin{table*}
\begin{center} \caption{NUV Flare Energies} \begin{tabular}{rcc}
\hline
\hline
Flare Star &Flare Energy (max)& Flare Energy (min)  \\
&(erg)&(erg)\\
\hline
\hline
Hyades-B  04:26:04.4 +17:07:14.0&4.5E+29&4.5E+29\\ Hyades-B  04:27:33.6 +16:52:22.0&3.9E+27&3.9E+27\\
Hyades-B  04:28:42.7 +17:11:50.3&1.3E+28&2.3E+27\\
Pleiades-(A)  03:42:35.6 +21:50:31.0&1.1E+29&2.7E+27\\
Pleiades-(B)  03:43:35.5 +26:21:31.1&2.1E+28&3.4E+27\\
Pleiades-(B)  03:43:35.5 +26:21:31.1&1.8E+28&2.8E+27\\
Pleiades-(B)  03:45:43.6 +26:05:05.0&4.9E+28&3.1E+27\\
\hline
\hline
\end{tabular}
\end{center}
\end{table*}

\newpage
\begin{figure*}
\center
{\plotone{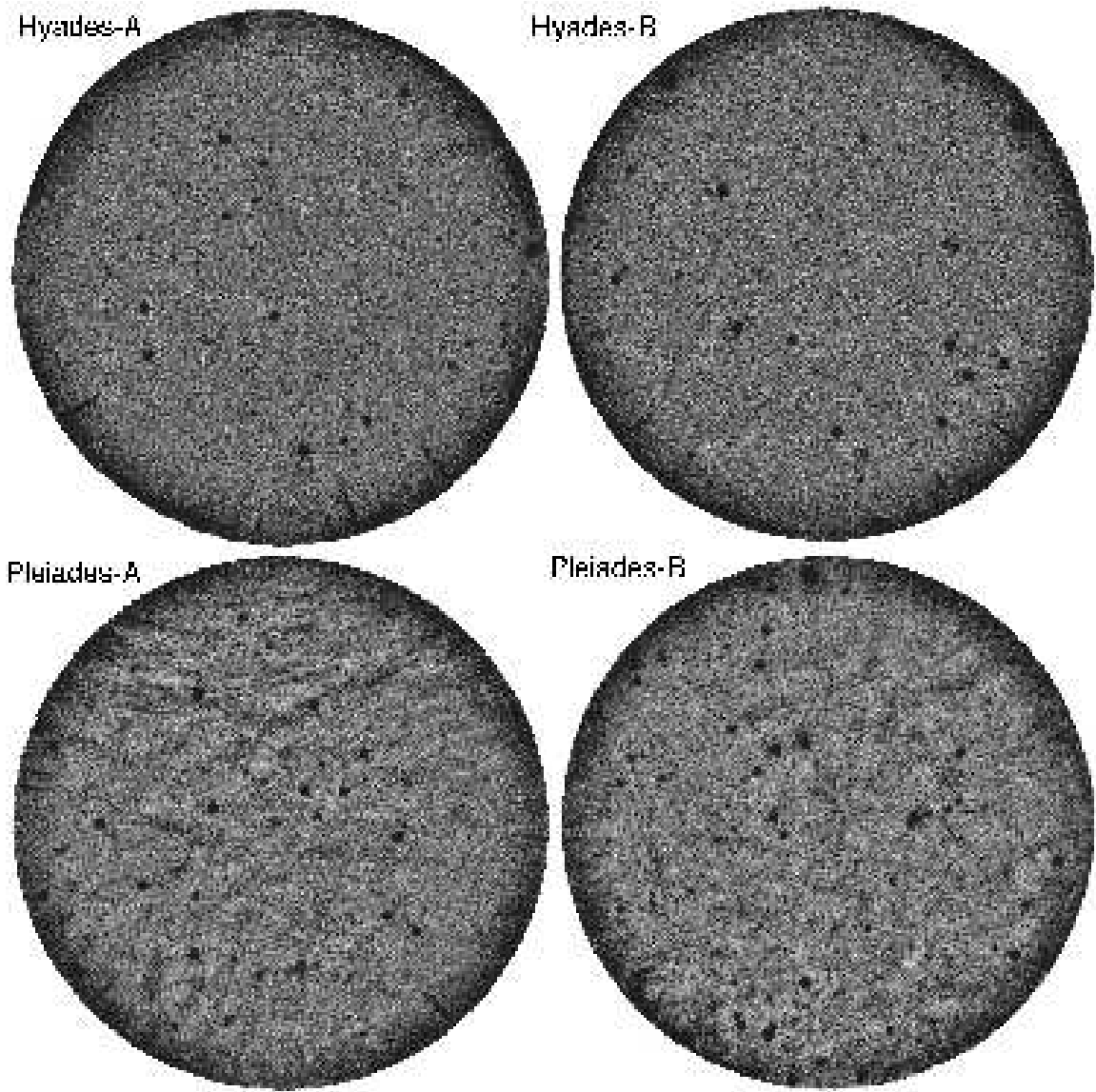}}
\caption{$\it GALEX$ NUV images of the Hyades A $\&$ B and Pleiades A $\&$ B fields scaled to the same photon count rate. The streaks on the lower edge of the both Hyades images are due to scattered/reflected light from UV bright objects just beyond the nominal fields of view. Note the complexity of NUV emission from interstellar and nebular gas and dust in both Pleiades images.}
\label{Figure 1}
\end{figure*}

\begin{figure*}
\center
{\plotone{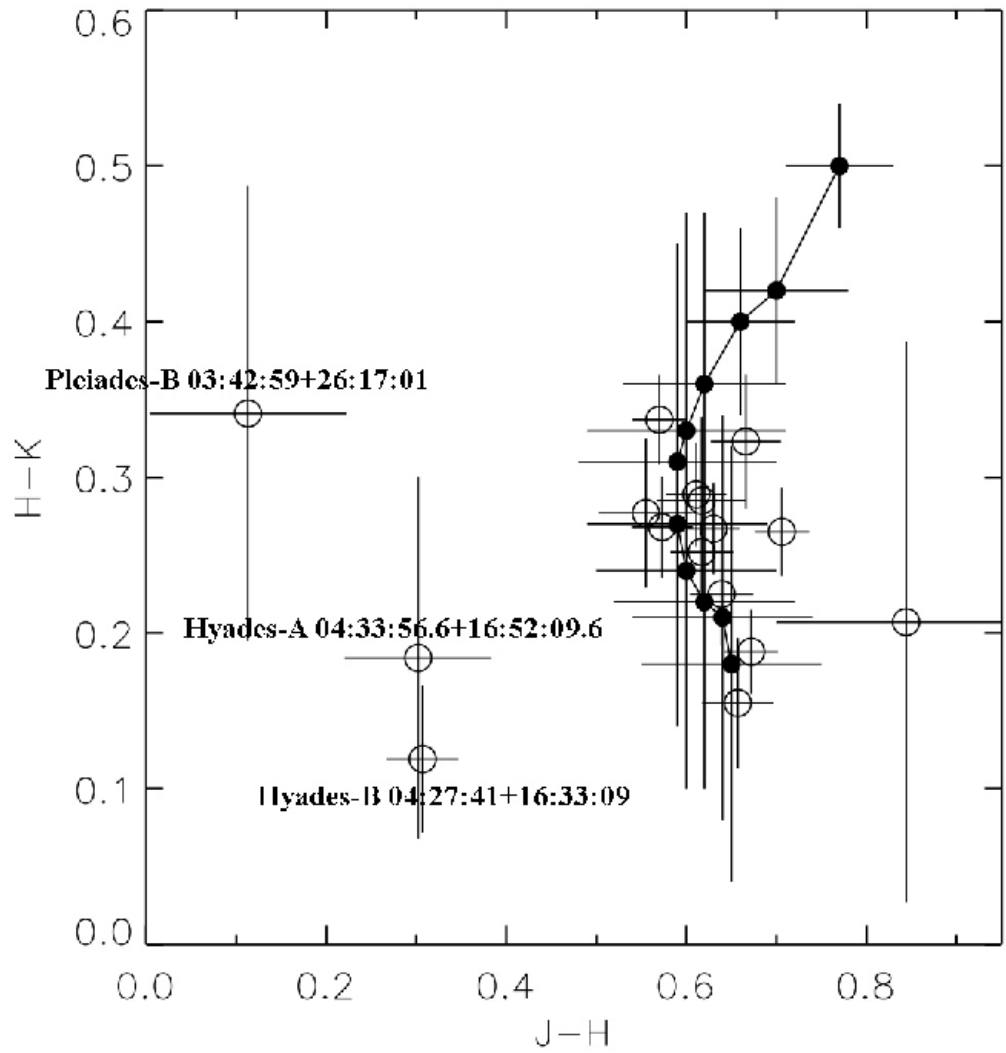}}
\caption{Plot of 2MASS (H - K) versus (J - H) color magnitudes for the 16 UV variable sources listed in Table 1 (open circles). Also shown is the locus of such color-color magnitudes for all 2MASS M0 to M9-type stars (filled circles) taken from West et al. (2008). Note that the majority of the UV variable sources lie within the color-color range for M-type stars. }
\label{Figure 2}
\end{figure*}

\begin{figure*}
\center
{\plotone{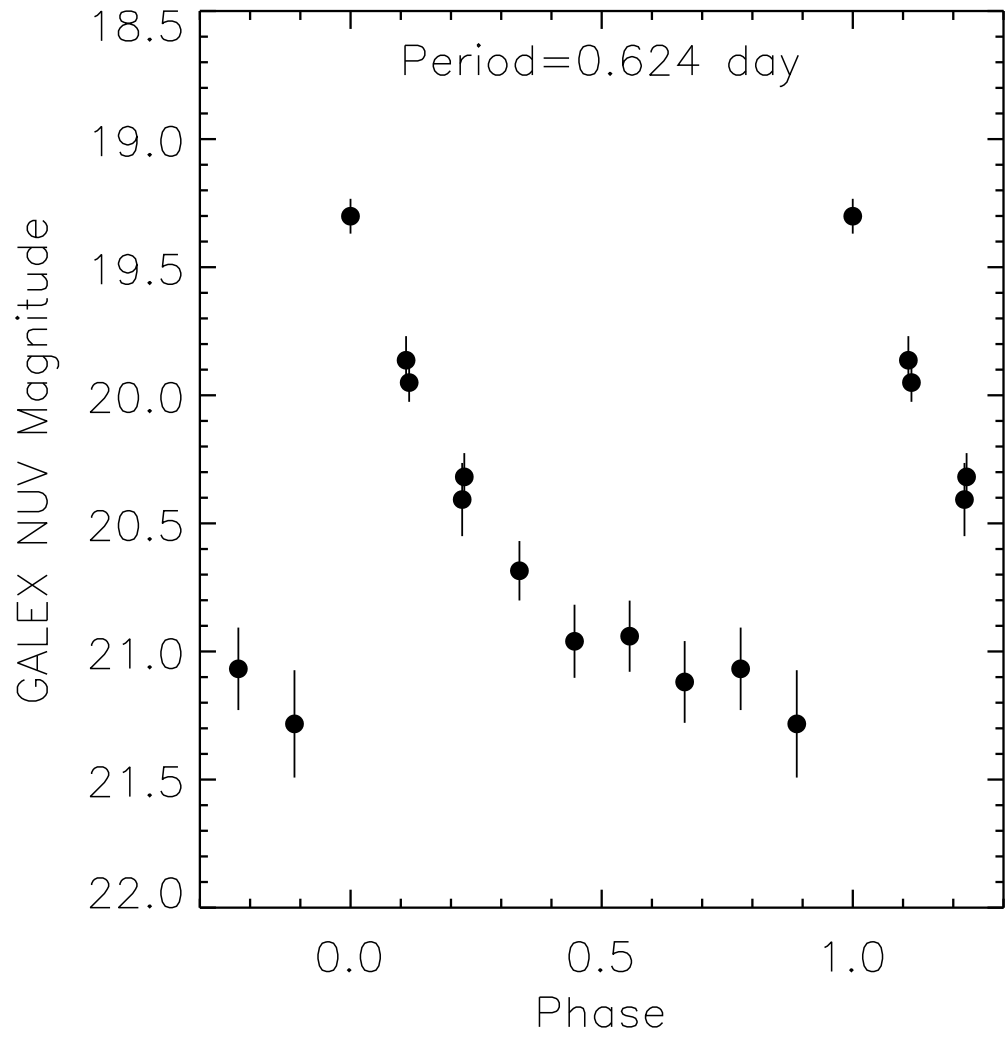}}
\caption{$\it GALEX$ NUV light-curve for the Hyades-B source 04:28:32.4 +16:58:21.6. The best fit period for this RR Lyrae star is 0.624 d. }
\label{Figure 3}
\end{figure*}

\begin{figure*}
\center
{\plotone{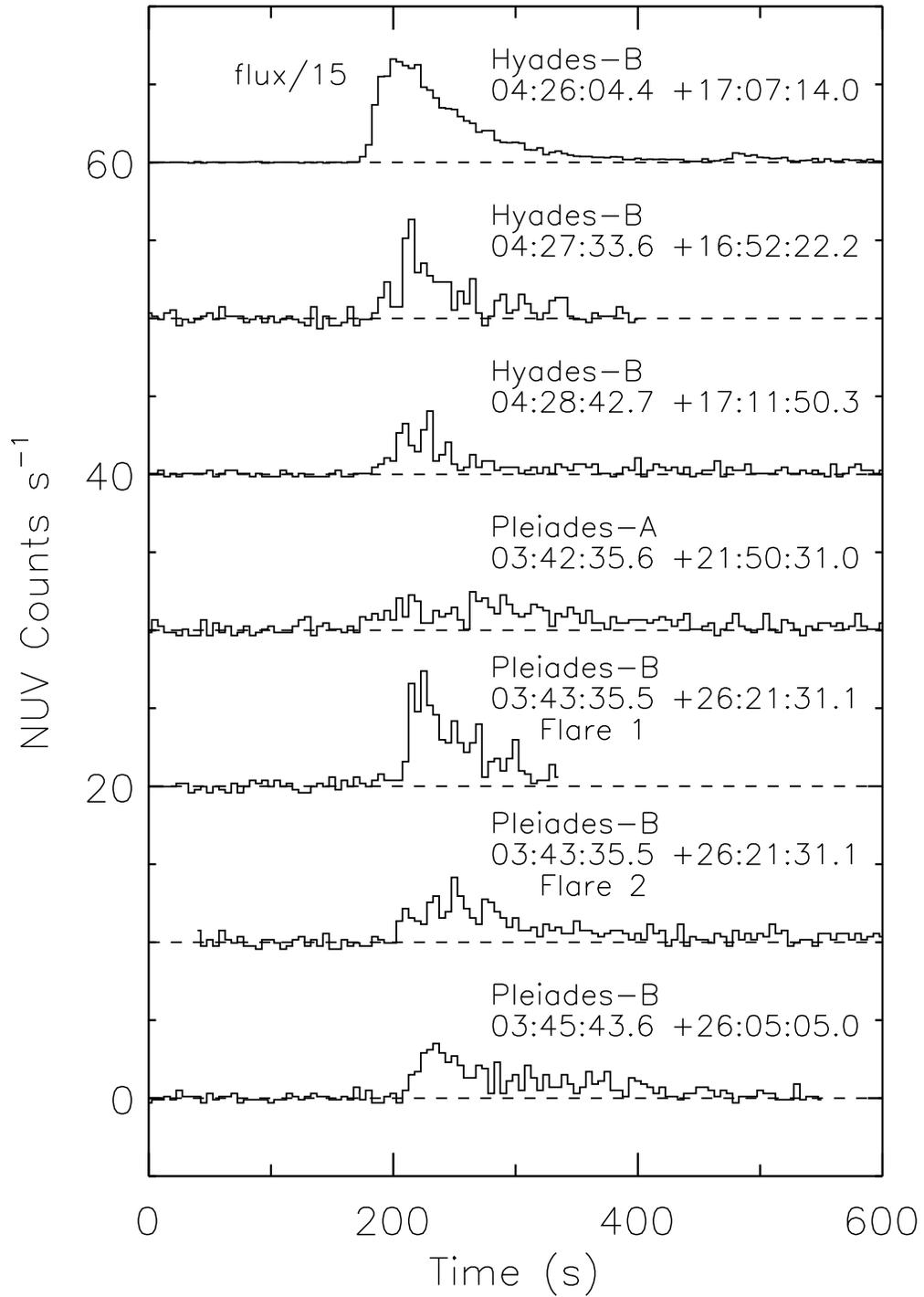}}
\caption{NUV light-curves (photon cts s$^{-1}$ versus time) for the 6 objects that exhibited flaring signatures during the $\it GALEX$ observations of the Hyades and Pleiades cluster fields.}
\label{Figure 4}
\end{figure*}

\begin{figure*}
\center
{\plotone{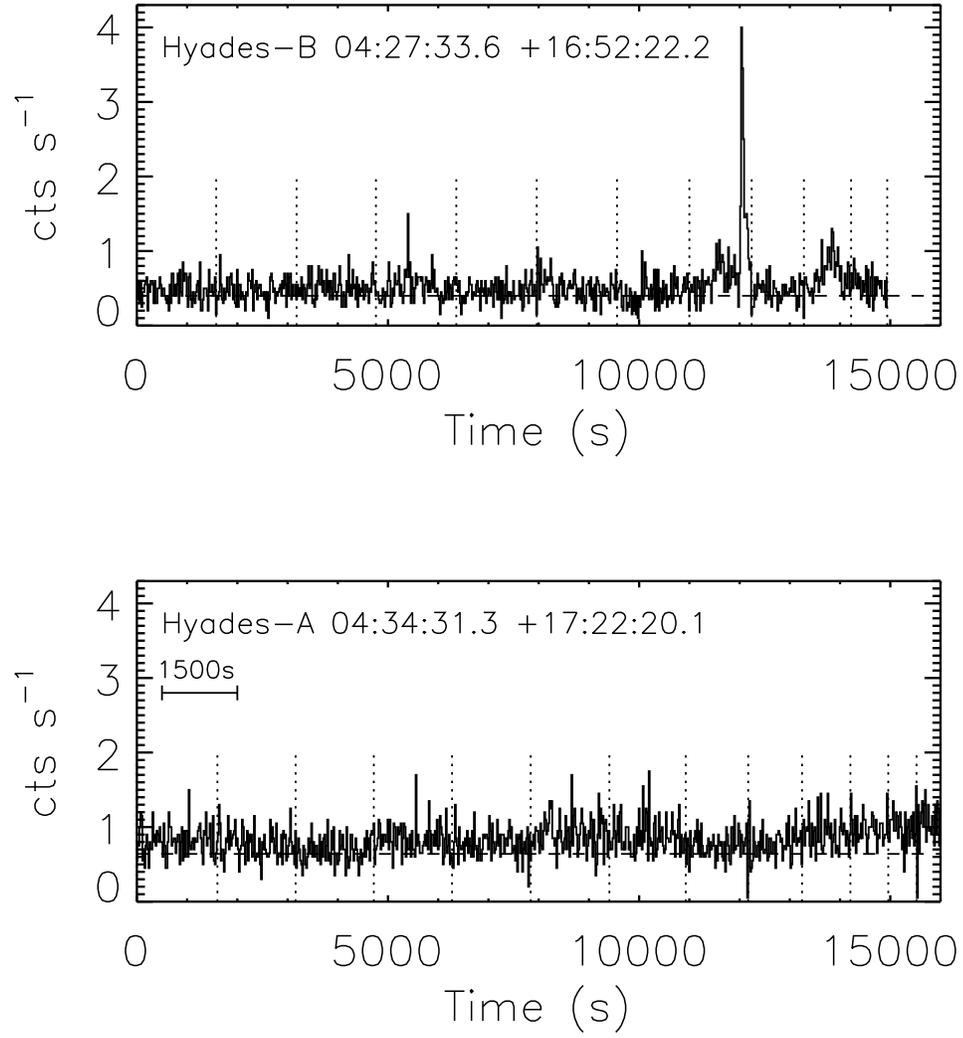}}
\caption{Concatenated series of 10 photon flux (cts s$^{-1}$) versus time light-curves (each of length $\sim$ 1500 s) for two
of the observed M-type sources. Significant gaps in time (due to satellite operational constraints) between each exposure are indicated by the dotted vertical lines. Note the large and far smaller short-term flare events on the upper plot. The source Hyades-A 04:34:31.3 +17:22:20.1 exhibits quasi-constant NUV emission for the majority of the exposure period, with a slowly increasing flux towards the last 3000 s of the exposures. }
\label{Figure 5}
\end{figure*}

\begin{figure*}
\center
{\plotone{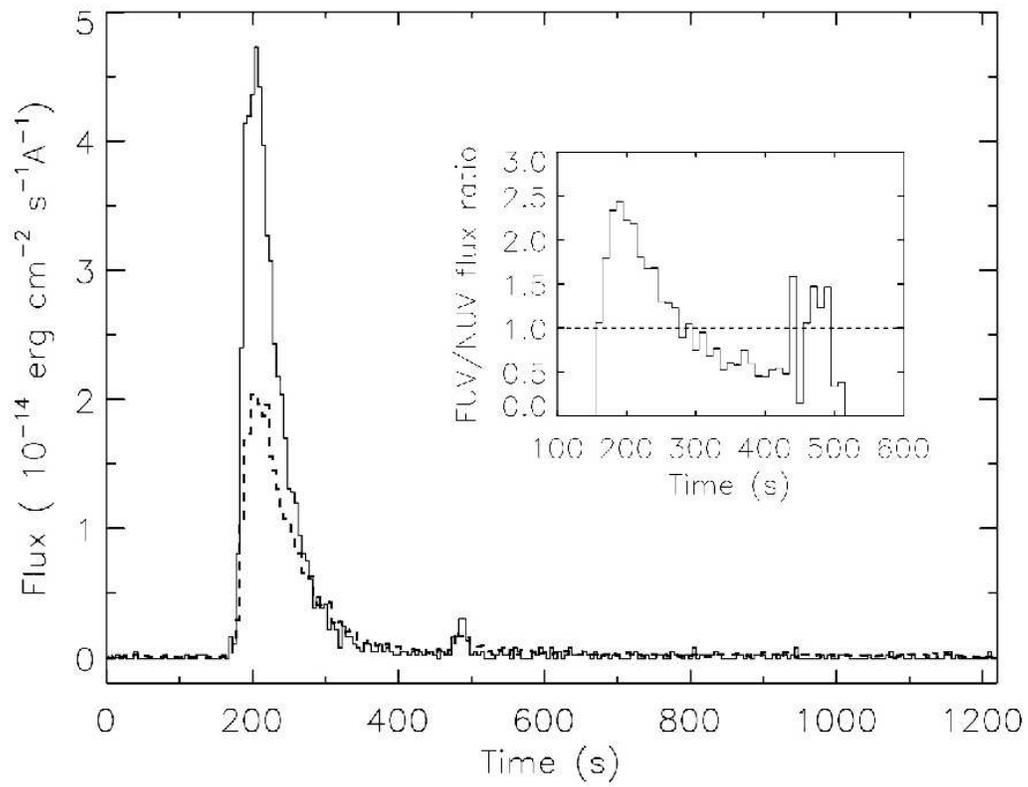}}
\caption{Comparison of the$\it GALEX$ FUV (full line) and NUV (dashed line) light-curves for the flare event observed on Hyades-B source 04:26:04.4 +17:07:14.0. Inserted is a plot of the FUV/NUV flux ratio as a function of time over the same exposure period.}
\label{Figure 6}
\end{figure*}


\begin{thebibliography}{}

\bibitem[2001]{adams01} Adams, J., Satuffer, J., Monet, D. et al., 2001, \aj, 121, 2053

\bibitem[Ayres et al. 2001]{ayres01} Ayres, T., Brown, A., Osten, R. et al., 2001, \apj, 549, 554-

\bibitem[1996]{bertin96} Bertin, E. and Arnouts, S., 1996, A $\&$ A, 117, 393

\bibitem[Chavushian $\&$ Gharibjanian 1975]{chav75} Chavushian, H. $\&$ Gharibjanian, A., 1975, Astrophysics, 11.565

\bibitem[Cutri et al. 2003]{cutri03} Cutri, R.M. et al., 2003, 2MASS All Sky Catalog of Point Sources (The IRSA 2MASS All-Sky Point Source catalog, NASA/IPAC Infrared Science Archive. http://irsa.ipac.caltech.edu/applications/Gator)

\bibitem[2004]{deac04} Deacon, N. $\&$ Hambly, N., 2004, \aap, 416, 125

\bibitem[Dobbie et al. 2002]{dobbie02} Dobbie, P., Kenyon, F., Jameson, R. et al., 2002, \mnras, 329, 543

\bibitem[Finlator et al. 2000]{fin00} Finlator, K., Ivezic, Z., Fan, X. et al., 2000, \aj, 120, 2615

\bibitem[Fuhrmeister et al. 2008]{fuhr08} Fuhrmeister,B., Liefke, C., Schmitt, J. $\&$ Reiners, A., 2008, \aap, 487, 293

\bibitem[2003]{gibson03} Gibson, S. and Nordsieck, K., 2003, \apj, 589, 347

\bibitem[Gudel et al. 2003]{gudel03} Gudel, M., Audard, M., Kashyap, V. et al., 2003, \apj, 582, 423

\bibitem[Gudel et al. 2004]{gudel04} Gudel, M., Audard, M., Reale, F. et al., 2004, A $\&$ A, 416, 713

\bibitem[Hambaryan et al. 1997]{hamb97} Hambaryan, V., Mirzoyan, L, Wichmann, R. et al., 1997, Astrophysics, 40, 354

\bibitem[Hambly et al. 1991]{hamb91} Hambly, N., Jameson, R. $\&$ Hawkins, M., 1991, \mnras, 253, 1

\bibitem[Haro et al. 1982]{haro82} Haro, G., Chavira, E. $\&$ Gonzalez, G., 1982, Bol. Inst. Tonantzintla, 3, 3

\bibitem[Hawley et al. 2003]{hawley03} Hawley, S., Allred, M., Johns-Krull,  C. et al., 2003, \apj, 597, 535

\bibitem[1988]{leg88} Leggett, S. and Hawkins, M., 1988, \mnras, 234, 1065

\bibitem[1994]{leg94} Leggett, S., Harris, H. and Dahn, C., 1994, \aj, 108, 944

\bibitem[Martin et al. 2005]{mar05} Martin, D.C., Fanson, J., Schiminovich, D., et al. 2005, \apj, 619, L1

\bibitem[Micela et al. 1996]{mic96} Micela, G., Sciortino, S., Kashyap, V. et al., 1996, \apjs, 102, 75

\bibitem[Mirzoyan et al.1994]{mirz94} Mirzoyan, L., Ambaryan, V. and Garibdzhanyan, A., 1994, Ap, 37, 297

\bibitem[Monet et al. 1998]{monet03} Monet, D., Levine, S., Casian, B. et al., 2003, \aj, 125, 984

\bibitem[Morrissey et al. 2005]{mor05} Morrissey, P., Sciminovich, D., Barlow, T. et al., \apj, 619, L7

\bibitem[Morrissey et al. 2007]{mor07} Morrissey, P., Conrow, T., Barlow, T. et al., 2007, \apjs, 173, 682

\bibitem[Osten et al. 2005]{osten05} Osten, R., Hawley, S., Allred, J. et al., 2005, \apj, 621, 398

\bibitem[Panagia et al. 1995]{panagia95} Panagia, P. $\&$ Andrews, A., 1995, \mnras, 277, 423

\bibitem[Perryman et al. 1998]{perry98} Perryman, Brown, A., Lebreton, Y. et al., 1998, A $\&$ A, 331, 81

\bibitem[Pillitteri et al. 2005]{pillitteri05} Pillitteri, I., Micela, G., Reale, F. and Sciortino, S., 2005, A $\&$ A, 430, 155

\bibitem[Reid 1992]{reid92} Reid, N., 1992, \mnras, 257, 257

\bibitem[Reid et al. 1995]{reid95} Reid, N., Hawley, S. and Mateo, M., 1995, \mnras, 272, 828

\bibitem[2005]{robinson05} Robinson, R., Wheatley, J., Welsh, B.Y. et al., 2005, \apj, 633, 447

\bibitem[Segura et al. 2005]{segura05} Segura, A., Kasting, J., Meadows, V. et al., 2005, AsBio, 5, 706

\bibitem[Soderblom et al. 2005]{sod05} Soderblom, D., nelan, E., Benedict, G. et al., \aj, 129, 1616

\bibitem[2008]{sol08} Sollima, A., Cacciari, C., Arkharov, A. et al., 2008, \mnras, 384, 1583

\bibitem[Stauffer et al. 1991]{stauf91} Stauffer, J., Klemola, A., Prosser, C. $\&$ Probst, R., 1991, \aj, 101, 980

\bibitem[Stauffer et al.1997]{stauffer97} Stauffer, J., Balachandran, S., Krishnamurthi, A. et al., 1997, \apj, 475, 604

\bibitem[Stauffer et al. 2007]{stauf07} Stauffer, J., Hartmann, L., Fazio, G. et al., 2007, \apjs, 172, 663

\bibitem[Stern et al. 1995]{stern95} Stern, R., Schmitt, J. and Kahabka, P., 1995, \apj, 448, 683

\bibitem[Terndrup et al. 2000]{terndrup00} Terndrup, D., Stauffer, J., Pinsonneault, M. et al., 2000, \aj, 119, 1303

\bibitem[Welsh et al. 2005]{welsh05} Welsh, B.Y., Wheatley, J., Heafield, K. et al., 2005, \aj, 130, 825

\bibitem[Welsh et al. 2007]{welsh07} Welsh, B.Y., Wheatley, J.M., Seibert, M. et al., 2007, \apjs, 173, 673

\bibitem[2004]{west04} West, A., Hawley, S., Walkowicz, L. et al., 2004, \aj, 128, 426

\bibitem[West et al. 2008]{west08} West, A., Hawley, S., Bochanski, J. et al., 2008, \aj, 135, 785

\bibitem[Wheatley et al. 2005]{wheat05} Wheatley, J.M., Welsh, B.Y., Siegmund, O. et al., 2005, \apj, 619, L123

\bibitem[Wheatley et al. 2008]{wheat08} Wheatley, J.M., Welsh, B.Y. and Browne, S.E., 2008, \aj, 136, 259

\bibitem[York et al. 2000]{york00} York,D. G. et al., 2000, \aj, 120, 1579

\end{thebibliography}
\end{document}